\documentclass[preprint]{aastex}
\usepackage{epsfig}
\begin{document}

\title{Discovery and Implications of a New Large-Scale Stellar Bar in NGC 5248}
\author{
Shardha Jogee$^{1}$, 
Johan  H. Knapen$^{2,3}$, 
Seppo  Laine$^{4}$, 
Isaac Shlosman$^{5,6}$, \\ Nick Z. Scoville$^{1}$, and Peter
Englmaier$^{7}$ 
}

\footnotetext[1]{
Division of Physics, Mathematics, and Astronomy, MS 105-24, 
California Institute of Technology, Pasadena, CA 91125}
\footnotetext[2]{
Isaac Newton Group of Telescopes, Apartado 321, Santa Cruz de La Palma, E-38700 Spain}
\footnotetext[3]{Department of Physical Sciences, University of 
Hertfordshire, Hatfield, Herts AL10 9AB, UK}
\footnotetext[4]{
Space Telescope Science Institute, 3700 San Martin
Drive, Baltimore, MD 21218}
\footnotetext[5]{Joint Institute for Laboratory Astrophysics, University of 
Colorado, Campus Box 440, Boulder, CO 80309-0440}
\footnotetext[6]{JILA Visiting Fellow. Permanent address: Department of Physics and Astronomy,
University of Kentucky, Lexington, KY 40506-0055}     
\footnotetext[7]{MPI f{\" u}r extraterrestrische Physik, Postfach 1312, 
85741 Garching, Germany}







\begin {abstract}
For decades, the grand-design SAB  spiral galaxy NGC 5248 
has been postulated to host a short bar 
of semi-major axis $22\arcsec$ (1.6 kpc).  
From  dynamical and morphological  arguments, we argue, however, 
that its   spiral structure  is being 
driven by a  large-scale bar whose corotation radius lies at 
$\sim 115\arcsec$ (8.6~kpc).
Our estimate is based partially on a deep $R$-band image 
which reveals that the feature previously thought to be 
an inclined disk is in fact an extended stellar bar.   
The bar is embedded within a fainter  outer disk  
visible out to a radius of  $230\arcsec$ (17.2 kpc).
The bar has a deprojected ellipticity of 0.44 and 
a semi-major axis of $95\arcsec$ (7.1~kpc).
The classical  grand-design spirals  of NGC 5248, 
prominent  in $B$, $R$, and $K$  light, 
lie on the leading edge of the large-scale stellar 
bar and are accompanied by  concave dust lanes 
out to at least $70\arcsec$. 
The offset between the dust and young stars is consistent 
with our understanding of gas flows in barred galaxies, 
where shocks along the leading edges of a  moderately strong  bar
compress the gas to form massive young stars.
While in many  strongly barred galaxies, optical spiral 
arms are prominent outside the bar but not within it, 
NGC 5248 illustrates how  intense star formation along a 
moderately strong bar can lead to conspicuous open spiral arms 
within the bar itself. NGC 5248 also  provides a clear example  of how  
a  large-scale stellar bar embedded within a faint outer 
optical disk can be misidentified as   an  inclined disk  
when  imaging studies lack the sensitivity to detect the actual 
outer disk. We discuss the implications for the 
estimated bar fraction at higher redshifts.

\end{abstract}

\section {Introduction}

Grand-design spiral galaxies  generally host two long symmetric 
spiral arms which  can be followed  over a large azimuthal 
angle and  dominate the optical disk, e.g., 
NGC~1300  (Sandage 1961), 
NGC~895, NGC~1566  (Elmegreen \& Elmegreen 1995), 
M51,  M100  (Elmegreen, Seiden, \& Elmegreen  1989) 
and NGC~5248  (e.g., Patsis, Grosbol, \& Hiotelis 1997). 
Within the framework of the spiral density wave (SDW) 
theory  (e.g., Lin \& Shu 1964)
spiral arms in disk galaxies 
result from a traveling   wave pattern  that remains 
quasi-stationary in a frame of reference rotating around the 
galaxy  center at a certain pattern speed. 
Spiral density waves can be excited  by a companion 
or by a bar. 
The spiral waves can get swing amplified as they
are sheared  from leading to trailing (e.g., Goldreich \& Lynden-Bell 1965) 
provided the disk is cold enough.

In this Letter, we focus on the driving mechanism for a  SDW in 
the nearby ($D$=15 Mpc) well studied  grand-design spiral 
NGC~5248 and discuss the implications of an extended stellar bar in 
this galaxy. 
Optical imaging  (e.g., Fig.~1)  of NGC~5248 reveals two bright,  
relatively symmetric 
spiral  arms from about $30\arcsec$ (2.2 kpc)  to  $80\arcsec$ (6.0 kpc). 
NGC~5248 does not show   evidence of any recent or ongoing 
interaction with its three faint irregular neighbors 
(UGC 8575, UGC 8614, and UGC 8629; Zaritsky et al. 1997) 
which have similar redshifts and  are  
located more than $30\arcmin$ (135 kpc) away in projection.
It is therefore likely that the grand-design spirals in 
NGC~5248 are driven by a massive large-scale bar,  
which is  verified in the present work.
The circumnuclear ring of H{\sc ii}  regions 
at a radius of $5\arcsec$ (375 pc) (e.g., Elmegreen et al. 1997)  
also suggests the presence of a bar since  rings  are commonly  
associated with the dynamical resonances of a bar 
(e.g., review by Buta \& Combes 1996).

NGC~5248 is classified as SAB(rs)bc (de Vaucouleurs et al. 1991; hereafter RC3)  
and has been postulated  to have a weak stellar bar 
with an  ellipticity of 0.32 and a semi-major axis  of  $22\arcsec$ 
(1.6 kpc)  at a position angle (PA)  of $110^\circ$  (e.g., Martin 1995). 
This oval feature is evident in optical images (e.g., Fig.~1). 
However, the extent and morphology of the 
grand-design spirals  in NGC 5248  
strongly suggests  that the bar driving these spirals 
must  be significantly longer  and more massive  than previously assumed. 
The two bright inner stellar  
spiral arms  have prominent  dust lanes on their 
\it  inner (concave) sides  \rm   (Fig.~1) 
out to a radius $r \ge 70\arcsec$ (5 kpc). 
Such a dust lane morphology is  expected to exist only 
inside the corotation resonance (CR)  of the arms, 
where  young stars  which form when  gas is compressed by 
shocks,   seen as dust lanes,  
can overtake the pattern  
From this argument alone, we expect that the 
CR  of the spiral pattern in NGC~5248 is  
beyond $70\arcsec$.
If the bar and spirals in this galaxy have the same pattern speed, 
the latter must be driven by a stellar bar whose semi-major axis is
comparable in size to that of their common CR. 
Large-scale bars are expected to end at or near CR based on the
studies of stellar orbits (Contopoulos \& Papayannopoulos 1980), 
and on the  analysis of  shapes of  offset dust lanes within 
bars (Athanassoula 1992). 
In order to test the existence  of an extended    bar in NGC~5248,  
we obtained a deep $R$-band image with a large field of view ($\S$ 2). 
While this Letter focuses on the observational aspects of 
the large-scale bar in NGC~5248 and  its  consequences for galactic
dynamics in nearby and  high redshift galaxies,  the subsequent paper
(Jogee et al. 2002, hereafter Paper II) addresses gas dynamics in the 
inner few kpc of this galaxy, based on  multi-wavelength observations 
and modeling.


\section {Observations and Data Reduction}

NGC~5248 was imaged through  the Harris $R$-band  filter for 
a total  of 30  minutes using the 
Wide Field Camera  on  the 2.5 m Isaac Newton Telescope (INT)  in La Palma 
in August 2001.  The data frames had a plate scale of 
0\farcs 33/pixel, a field of view (fov) of 11\farcm 3 $\times$  22\farcm 5, 
and an average seeing of 1\farcs 8.
Photometric calibration was performed by comparison with a shorter image
of the galaxy in the same band, taken on the same night with the 1 m
Jacobus Kapteyn Telescope. 
The $R$-band   surface brightness (SB)  goes down to 26.5 mag arcsec$^{-2}$. 
Bias-subtracted and  flat-fielded frames were obtained from 
the INT data reduction pipeline. Sky-fringing at the  1 \% level is present,  
an effect known for the  thinned EEV  chip CCD4. 
We used the IRAF package  for 
fixing bad pixels, cleaning cosmic rays,  and 
combining frames   into a final image. 
After masking out stars and faint background galaxies, 
isophotal analysis 
was performed with the  ``isophote'' package in IRAF. 
We also refer to a $K'$-image obtained in May 2000 
using the INGRID camera on the 4.2~m William Herschel Telescope. This 
image, and other data described in this paper, are fully detailed in 
Paper~II.

\section {The extended bar and faint outer disk}

The  $R$-band image  (Fig.~1) reveals a hitherto unknown 
large-scale stellar bar-like feature embedded within a fainter,  
more circular,  outer 
disk  which is visible out to a radius of  $230\arcsec$ (17.2 kpc).  
Below we describe the morphological, photometric,  and  
dynamical properties of this bar. 
Earlier  optical and NIR images  lacked the   sensitivity and/or field of 
view  to clearly detect the end of the 
large-scale stellar bar and the  faint outer disk.  
As a result,  the  bar  was 
misinterpreted as  an inclined outer disk. 
The new  $R$-band image  (Fig. 1)  shows the actual 
outer disk and  the full extent of the  bar. 
In order to quantitatively describe the features 
in NGC~5248, we performed an isophotal analysis (Fig.~2) 
of the  $R$-band and available  $K$-band  images.   
To minimize the effects of extinction, the $K$-band  image of 
$0 \farcs 6$ seeing  is used  out to a radius of  $50\arcsec$. 
Beyond this point, its S/N ratio is low and we use 
the $R$-band  image.


The bright optical stellar spiral arms 
extend along the leading ($\S$ 4) edge of the bar  and 
cross its major axis (PA $\sim 135^\circ$) at  around $95\arcsec$ 
(Fig. 1). This  suggests that the bar has a 
semi-major axis of about $95\arcsec$  (7.1 kpc).
Large-scale bars  typically tend to end  near the  ultra harmonic resonance 
  where chaos and orbital instability set in. 
The ratio of corotation  to bar radii ($R_{\rm CR}$/$R_{\rm bar}$) 
is expected to lie in the range of $1.2\pm 0.2$ as found empirically 
by Athanassoula (1992). This corresponds to a CR 
of $\sim 115\arcsec$  (8.6~kpc) in NGC~5248 -- a value supported by 
photometric evidence.
As seen in Fig.~2, between $60\arcsec$ to $115\arcsec$, the projected
ellipticity {\it e} rises to a maximum of 0.52  and the
PA settles to  a fairly constant value of  $135^\circ$, 
after some initial variation between $120^\circ$--$135^\circ$.
This small variation is due to the unusual prominence of the spiral arms 
within the bar. The peak in $e$  over a  fairly constant  PA, 
as observed in NGC~5248, is the behavior expected of bars  
(e.g., Knapen et al. 1995). 


Between $a$~=~$115\arcsec$--$150\arcsec$, secondary faint spirals 
which stem off the bar at lower radii become particularly prominent.
They dominate the fitted isophotes which show a varying \it e \rm  
of 0.41--0.46  (Fig.~2).  Such features are reminiscent of ``plumes''
which are seen in other barred systems such as NGC~1433 and 
may be secondary compression zones  which arise near CR (Buta 1984).
Between  $150\arcsec$--$230\arcsec$,  the isophotes are dominated by 
the outer disk and  very faint spirals  which extend away from 
the bar and cover $90^\circ$--$100^\circ$ in azimuth (Fig. 1). 
This leads to a low but varying $e$ (Fig. 2).
The outer Lindlad resonance (OLR), where arms are expected to end,
is probably around $230 \arcsec$.  
The PA of the projected disk  is  relatively constant at $105^\circ$. 
Assuming the outer disk is intrinsically circular, we derive 
a PA of $105^\circ \pm 2^\circ$ for the line of nodes  and an inclination  
of  $40^\circ \pm 4^\circ$.  With these parameters, we performed 
a two-dimensional deprojection of the $R$-band   and $K$-band   images.
The deprojected  disk  (Paper~II) is  circular  with 
\it e \rm  below 0.02  (Fig.~2) 
while the deprojected  bar is moderately strong with 
a peak  \it e  \rm of 0.44. 
We note that the deprojection  is somewhat uncertain because 
the intrinsic shape of the disk  between CR and OLR 
is expected to be slightly elongated along the bar minor axis.

We now turn our attention to the inner  regions ($r \le 50\arcsec$) 
of NGC~5248.   
A very weak oval feature  is present in  the inner  $3\arcsec$ 
radius of the $K$-band image (Fig.~3) where the  deprojected  \it e \rm 
varies from 0.18 to 0.10  (Fig.~2).  
Two  $K$-band spirals,  extending from this oval, are  
lined with ``hot spots''  between  $3\arcsec$--$9\arcsec$.  They cross 
the optically-visible 
starburst ring at $\sim 5\arcsec$ and between $5\arcsec$--$8\arcsec$, 
they delineate  the bright super star clusters 
seen in  {\it HST} images (see Paper~II). 
Near the  bar major axis, the spirals fade out, and  \it e \rm  
falls from 0.20 to 0.11  between $10\arcsec$--$13\arcsec$ (Fig.~2). 
Between $17\arcsec$--$26\arcsec$,  the stellar distribution 
looks smoother  (Fig.~3) and the  $K$-band spirals fade as they approach
the bar major axis. 
Further out,  between $30\arcsec$--$50\arcsec$, the 
bright patchy outer $K$-band spirals (Fig.~3)  dominate, 
leading to a higher $e$ (0.15--0.30) and  a smoothly varying PA. 
The  oval-shaped  feature of radius $\sim$~$22\arcsec$  
which is visible in the projected images  (e.g., Fig.~1) 
has been interpreted as a 
primary bar of semi-major axis $22\arcsec$  (1.6 kpc) at a 
PA   of $110^\circ$ in  earlier studies (e.g., Martin 1995). 
However,  after deprojection, this feature has a  
peak $e$ of only 0.15 (Fig.~2). 
Therefore, the oval appearance of this  feature in the sky plane  
seems to be largely due to projection effects. 
The dip in $e$ at around $26\arcsec$ is due to the K-band 
arms joining the oval at its minor axis. 
Uncertainties in the deprojection are unlikely to change this 
conclusion, namely, there appears to be no signs of a bar of size $22\arcsec$ 
in NGC~5248.



\section {Discussion: Implications for barred spirals locally and at high redshifts }

NGC~5248 shows a grand-design spiral  morphology in stars, dust, and gas, 
as  traced by  $B$-band  (Paper~II),  $R$-band (Fig.~1),  
$K$-band (Fig.~3), $B-I$, H$\alpha$,  and CO maps (Paper~II).
In $B$,  $R$ and $K$ light, there are prominent 
stellar spiral arms  between about  $30\arcsec$--$80\arcsec$ (2.2--6.0 kpc), 
on the  leading edge of the large-scale stellar bar. 
The stellar arms 
have prominent dust spirals on their  inner (concave) sides 
out to at least  $70\arcsec$ (e.g., Fig.~1), as is expected 
inside the CR. 
The observed  offset between  the dust  and stellar spirals  
is consistent  with   our understanding of gas flows in 
barred galaxies  where   offset dust lanes along the leading edges 
of the bar delineate the shocks which compress the gas, 
leading to the formation  of  massive young stars. 
Within this framework, the bar is rotating clockwise, and 
the patchy  $B$, $R$, and $K$ stellar spiral arms  immediately offset from  
the dust lanes are  made of  \it young \rm stars. 
It is particularly noteworthy that the appearance of the $K$-band spirals 
in NGC~5248 is dominated by young stars.  $K$-band light 
is generally assumed to trace  old stellar populations, 
but it can be dominated by very young (8-10 Myr)  
supergiants in starburst regions (e.g., Knapen et al. 1995) 
Along the $K$-band arms in NGC~5248, there is  good 
spatial coincidence between peaks in  $K$-band emission 
and H{\sc ii}  regions  as traced by  H$\alpha$ images. 
The patchy morphology of the $K$-band spirals  is 
in stark contrast to  the smooth appearance of old  stellar arms 
(e.g., Kennicutt \& Edgar 1986). Taken together, these properties 
suggest that young stars dominate the $K$-band spirals.

While grand-design optical spiral arms in many 
\it strongly \rm barred galaxies stem from the ends of a stellar bar, 
in NGC~5248,  open spiral arms appear conspicuous within the 
\it moderately  strong \rm bar itself, delineating intense SF.  
The SF  along the offset bar dust lanes 
in NGC~5248 is consistent with current understanding of gas flows and 
SF in barred galaxies.
In strong bars which exhibit almost straight offset dust lanes 
delineating shocks, little SF  is seen 
along the bar. Common examples are  NGC~1300 and  NGC~5383 (Tubbs 1982). 
Star formation is believed to be inhibited because of 
the strong shear in the postshock flow (e.g., Athanassoula 1992). 
In weaker bars, where  the offset dust lanes are curved, 
the weaker shocks and shear can induce SF rather than inhibit it. 
For instance,  the collapse of gas cores to form stars can be 
induced by weak shocks with speeds of order 20 to 30 
km~s$^{-1}$  (e.g., Vanhala \& Cameron 1998). 
It is thus common to see SF along weak bars, e.g., 
in M100 (Elmegreen et al. 1989), 
NGC~4254, and NGC~4303  (Koopmann 1997). 
NGC~5248 falls in this category as its large-scale bar  is 
only moderately strong with  curved 
offset dust lanes (Fig.~1), and a  deprojected  ellipticity  of 0.44. 

We note  that there exists a class of  weakly barred galaxies  
where bright optical spiral arms  appear to emerge from 
a relatively short oval bar-like feature and extend to twice 
the radius of this bar (Elmegreen \& Elmegreen 1995). 
We would have put NGC~5248 in this category 
if we had taken its primary stellar bar to be the oval 
feature of radius $22\arcsec$, as was long assumed 
to be the case.  However, with our identification 
of  the  large-scale  bar  of radius $95\arcsec$, 
the  bright  optical spiral arms with concave dust lanes 
in NGC~5248 turn out to lie inside the bar. The short
oval turns out to be primarily  the result of projection effects 
($\S$ 3). It is therefore imperative to ask whether galaxies 
with  bright extended  optical spiral arms apparently driven 
by a short oval feature,  are in fact hiding 
a more extended large-scale  bar.   
NGC~1566 and ESO 111-110 (Buta 1995) are two such possible 
candidates.

While the majority of nearby spiral galaxies are barred (e.g., 
Grosbol 2002),  a  low fraction of  bars ($<$ 10 \%) 
has been reported  in galaxies at redshifts $z \sim$ 0.5--0.8 (e.g., 
Abraham et al.  1999).   In this latter study, the authors apply the 
same bar identification method to nearby galaxies and 
to  spirals at  $z \sim$ 0.5--0.8. 
Images which trace  the rest-frame $B$-band light are used.
The inclination  $i$ is derived from an outer isophote which corresponds 
to an arbitrary fraction  (1\%)  of the peak flux. 
After removing the projected  ellipticity corresponding to $i$, 
a bar is identified based on the residual ellipticity  in the inner bright 
regions of the spiral galaxy. The authors find a much lower bar fraction 
at intermediate redshifts than locally. 
It has been suggested that this result  may be biased due to the  small 
sample size or  that bars are truly deficient 
at intermediate redshifts because of  dynamically hotter disks or  
an  enhanced  efficiency in bar destruction  (Abraham et al. 1999).  
Another possibility is that the  method used  is less 
efficient in identifying  bars at intermediate redshifts than locally. 
This  method relies  on the $B$-band surface brightness ratio ($R_{\rm 1}$) 
between the central regions and the outer disk beyond the bar end. 
In a barred spiral with  a bright central region and  a faint 
outer disk beyond the bar end,  $R_{\rm 1}$ is low,  and this method 
can readily fail: the outer 1\% isophote would lie inside the bar rather 
than in the faint, more circular outer disk and,  consequently, 
the bar would be misinterpreted as an inclined disk. 
Bars may be missed more frequently at  $z \sim$ 0.5--0.8 than 
locally with this method because of two factors: evolution and imaging depth. 
If there is evolution in the surface-brightness profiles of spirals 
with redshift   such that  $R_{\rm 1}$ is on average lower at higher 
redshifts, then bars may be missed more frequently at  $z \sim$ 0.5--0.8 
than locally. Detailed future studies with large samples 
are needed to quantify $R_{\rm 1}$,  but our suggestion is not ruled 
out by existing data. Spiral galaxies at intermediate redshifts have 
bright blue central colors indicative of recent starbursts 
(Ellis, Abraham, \& Dickinson 2001), and the outer disk/bar 
which have presumably formed recently may still be quite faint.
Another factor is that images of spirals at $z \sim$ 0.5--0.8 often 
do not reach the  same sensitivity level as images of nearby galaxies.
%

We also discuss  why a higher  bar fraction at 
$z \sim$ 0.5--0.8  is quite likely.  
It is well known that bars  or $m$=2 modes can be easily excited in 
a  dynamically cold disk  either spontaneously 
or  during tidal interactions and minor mergers. 
On the other hand,  major mergers tend to  produce highly disturbed 
systems  which  eventually evolve into  elliptical-like systems (e.g., 
Barnes 1992) through  violent relaxation. 
The fraction of spiral galaxies in pairs 
increases  with redshift out  to   $z \sim $  0.75--1 (Le  F\`evre  et al. 
2000), suggesting  enhanced interaction between spiral galaxies.
These interactions do not appear to involve major mergers since 
\it HST \rm  studies  hint  that  sizes  of spiral  galaxies  
(e.g., Lilly et al. 1998) 
and the  number counts of spirals   with $I<$22 mag  
in \it HST \rm  fields  (e.g., Glazebrook et al. 1995) 
do not  show   much evolution  out to   $z \sim $ 1.
Taken together, these results suggest  that conditions 
out to  $z \sim $ 1 are favorable to the formation of bars 
in disk galaxies. 
Furthermore, if the recently  discovered massive disk  
galaxy at $z$ =1.34 (Van Dokkum \& Stanford 2001) 
is representative, it reveals  that massive, organized, 
disks with ongoing SF  are already in place by 
$z$ =1. The   ring of  SF at a  $\sim$~10 kpc radius 
may be associated with the dynamical resonances of a bar  or 
alternatively could be an out-of-the-plane ring. 
%

\acknowledgments
Support for this  work was provided by 
a grant from the K. T. and E. L. Norris Foundation, and an 
NSF grant AST 99-81546 for S. J. and N. Z. S.,  and grants 
NAG 5-10823, HST GO-08123.01-97A, and WKU 516140-02-07 for I. S. 
We thank Thomas Augusteijn for INT service observations and 
Ron Buta for useful comments.
Data were  retrieved from the Isaac Newton Group (ING) archive. 
The William Herschel, Isaac Newton, and Jacobus Kapteyn Telescopes 
are  operated on  La 
Palma by the ING in the Spanish Observatorio del Roque de los Muchachos 
of the  Instituto de Astrof\'\i sica de Canarias.

\clearpage
\setcounter{figure}{0}
\begin{center} 
\epsfig{file=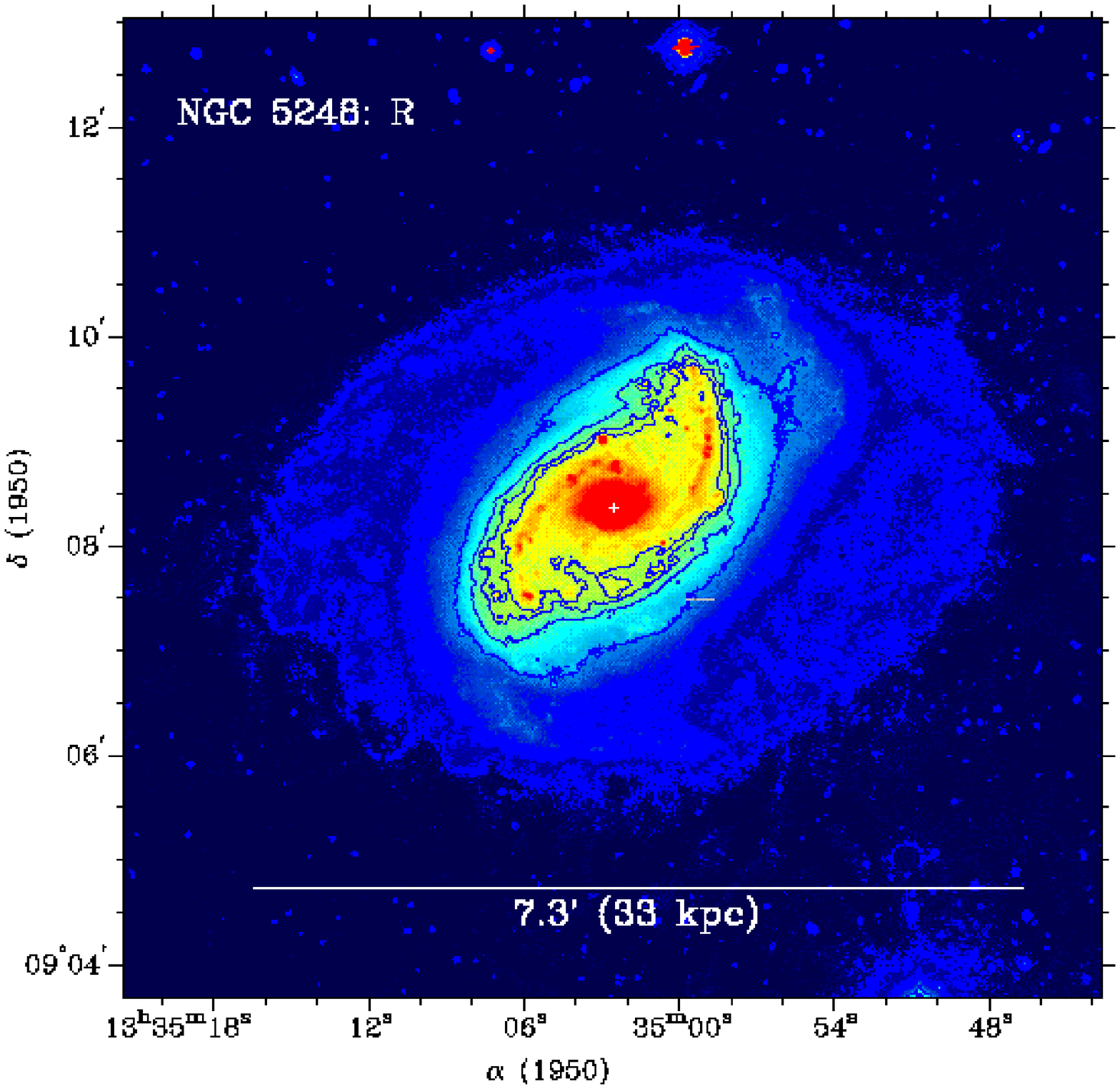,height=17.5cm}
\figcaption{
\rm 
The   $R$-band  image  
reveals a  hitherto unknown  prominent  stellar bar of size $95\arcsec$, 
embedded within a faint, more circular,  outer disk which is  visible 
out to a radius of  230\arcsec (17.2 kpc). 
The two  bright inner spirals  between $30\arcsec$ (2.2 kpc)  
and  $80\arcsec$   (6.0 kpc)  lie on the leading edge of the bar.
For a high resolution version of this figure, see 
ftp://ftp.astro.caltech.edu/users/sj/astroph/n5248-p1-highres.ps.gz
\label{fig1}}
\end{center} 

\clearpage
\setcounter{figure}{1}
\begin{center} 
\epsfig{file=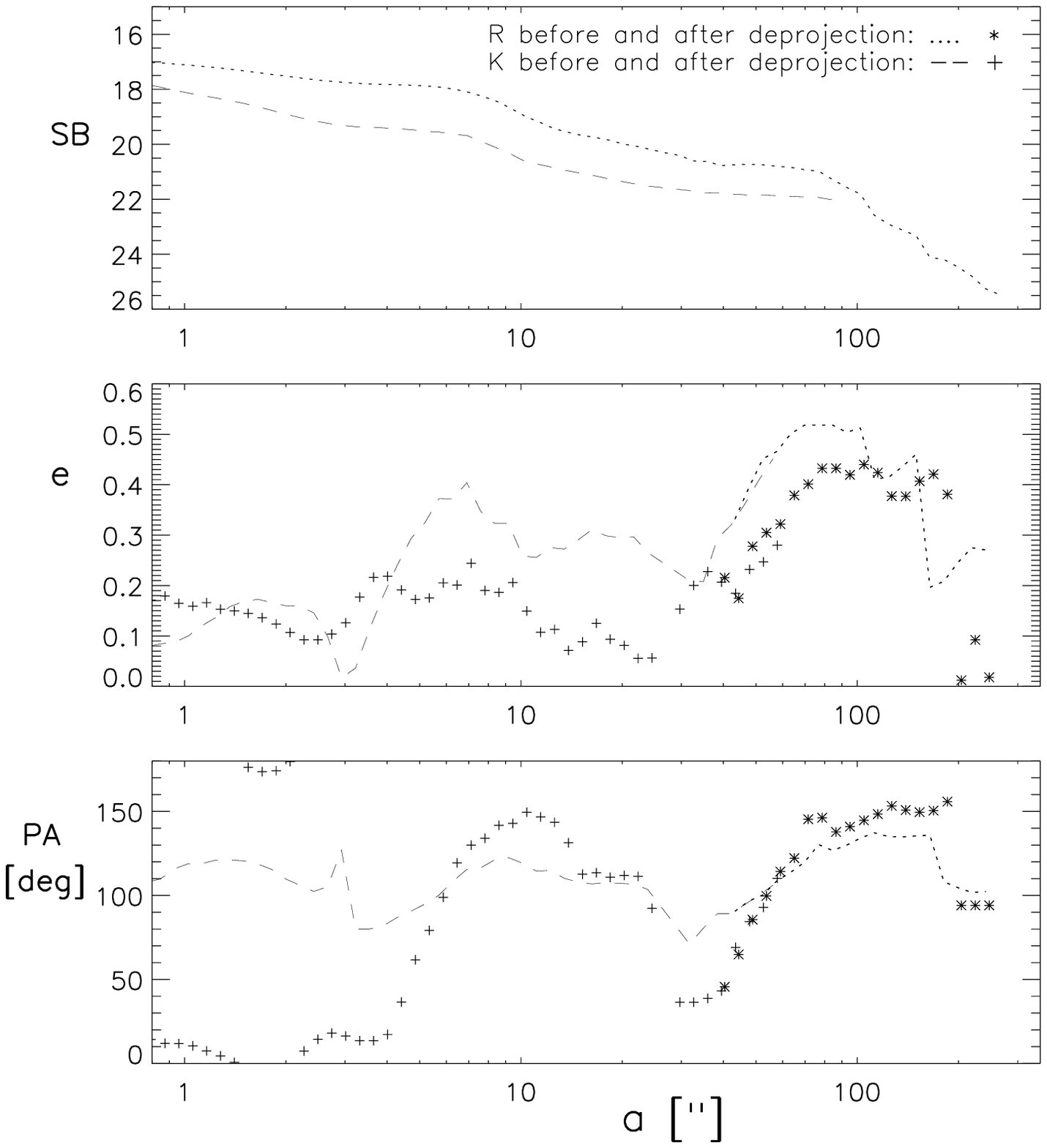}
\figcaption{
The radial profiles of surface brightness (SB), ellipticity ($e$),  
and  position angle (PA) of the azimuthally-averaged $K$-band and 
$R$-band  light are shown.
The dotted and dashed lines refer to the values before deprojection.
The symbols  refer to  deprojected values.
The $K$-band SB has been scaled by 1.32 so that
it can be plotted on the same scale as the  $R$-band profile.
In  deprojected images, we adopt the convention that PAs 
are measured anticlockwise from ``North''.
The deprojected  bar has an  ellipticity of 0.44.
\label{fig2}}
\end{center} 


\clearpage
\setcounter{figure}{2}
\begin{center} 
\vspace{-10.0cm}
\epsfig{file=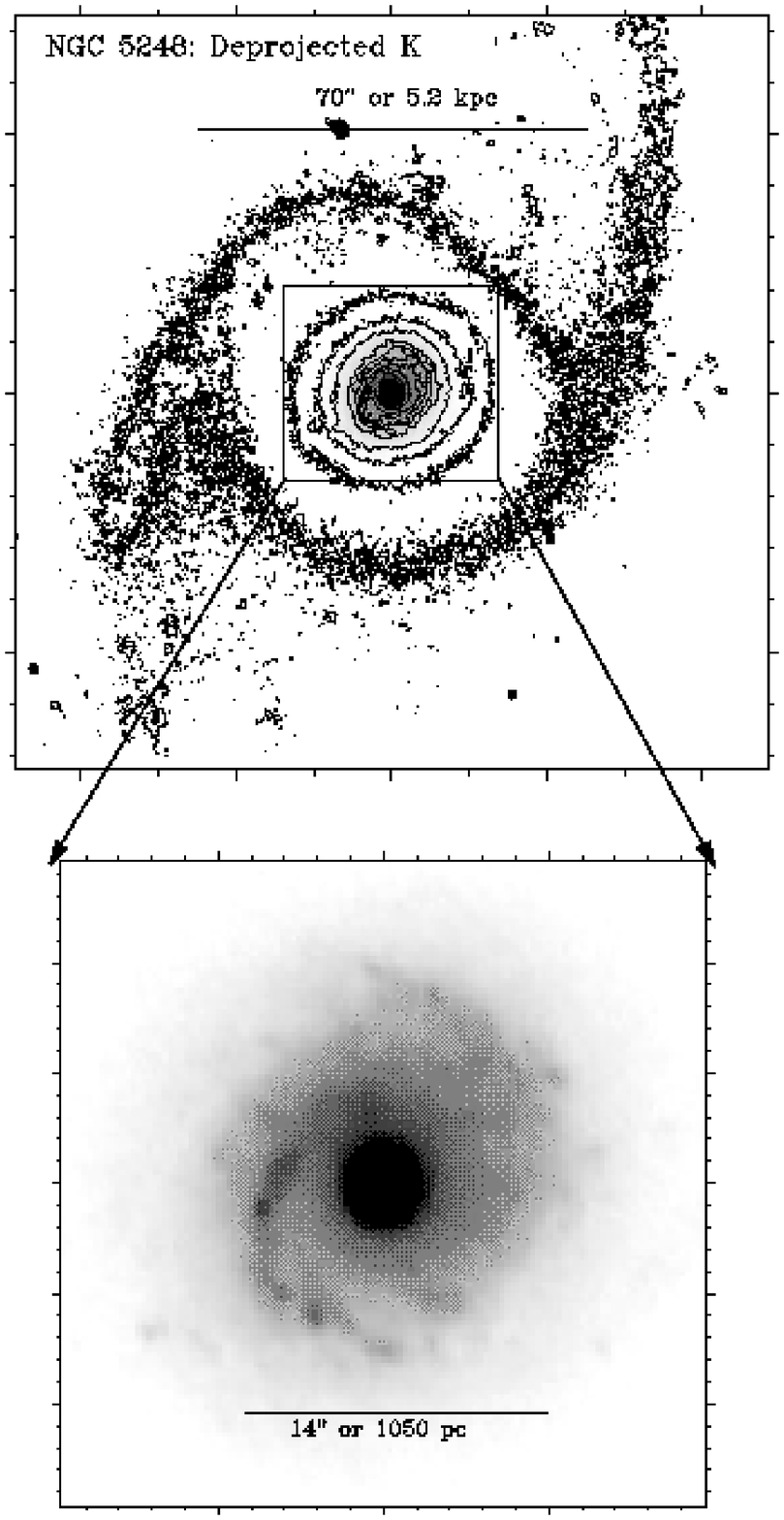,height=19cm}
\vspace{-1.6cm}
\figcaption{
The bottom and top panels show, respectively,  
the  central  $30 \arcsec$  (2.2 kpc) and  $2.4 \arcmin$ (10.8 kpc)
of the  deprojected $K$-band image. 
Notice 
the $K$-band spirals lined with ``hot'' spots  between 
$4\arcsec$--$9\arcsec$, the relatively axisymmetric distribution 
between $17\arcsec$--$26\arcsec$, and the patchy $K$-band   spirals 
beyond $30 \arcsec$. For a high resolution version of this figure, see 
ftp://ftp.astro.caltech.edu/users/sj/astroph/n5248-p1-highres.ps.gz
\label{fig3}}
\end{center} 
\end{document}